\documentclass{llncs}

\usepackage[toc,page]{appendix}
\usepackage{amsmath}
\usepackage{comment}
\usepackage{subfigure}

\usepackage{listings}
\usepackage{xcolor}
\usepackage{verbatim}
\usepackage{multirow}
\usepackage{graphicx}
\usepackage{url}
\usepackage{wasysym}
\usepackage{textcomp}
\usepackage{amssymb}
\usepackage{cite}
\usepackage{balance}
\usepackage{booktabs}
\usepackage{longtable}

\usepackage{array}
\usepackage{rotating}
\newcommand{\PreserveBackslash}[1]{\let\temp=\\#1\let\\=\temp}
\newcolumntype{C}[1]{>{\PreserveBackslash\centering}p{#1}}
\newcolumntype{R}[1]{>{\PreserveBackslash\raggedleft}p{#1}}
\newcolumntype{L}[1]{>{\PreserveBackslash\raggedright}p{#1}}

\usepackage[linesnumbered,lined, noend, algoruled,norelsize]{algorithm2e}
\usepackage{pifont}
\usepackage{mathtools}
\usepackage[colorlinks,linkcolor=blue,anchorcolor=blue,citecolor=blue]{hyperref}
\usepackage{flushend}
\usepackage{ulem}
\graphicspath{{./Figs/}}
\usepackage{algorithm2e}
\usepackage{seqsplit}
\SetAlFnt{\small}
\SetAlCapFnt{\small}
\SetAlCapNameFnt{\small}
\usepackage{indentfirst}

\begin{document}
\title{Characterizing Erasable Accounts in Ethereum}

\author{Xiaoqi Li\inst{1} \and Ting Chen\inst{2} \and Xiapu Luo\inst{1}\thanks{The corresponding author} \and Jiangshan Yu\inst{3}}

\institute{
	Department of Computing, The Hong Kong Polytechnic University, China.\\
	\email{csxqli@gmail.com, csxluo@comp.polyu.edu.hk}	
	\and
	Center for Cybersecurity, University of Electronic Science and Technology of China, China. \email{brokendragon@uestc.edu.cn}
	\and
	Faculty of Information Technology, Monash University, Australia.\\
	\email{jiangshan.yu@monash.edu}
    }

\maketitle

\begin{abstract}
\label{abs}
Being the most popular permissionless blockchain that supports smart contracts, Ethereum allows any user to create accounts on it. However, not all accounts matter. For example, the accounts due to attacks can be removed. In this paper, we conduct the first investigation on erasable accounts that can be removed to save system resources and even users' money (i.e., ETH or gas). In particular, we propose and develop a novel tool named \textsc{Glaser}, which analyzes the State DataBase of Ethereum to discover five kinds of erasable accounts. The experimental results show that \textsc{Glaser} can accurately reveal 508,482 erasable accounts and these accounts lead to users wasting more than 106 million dollars. \textsc{Glaser} can help stop further economic loss caused by these detected accounts. Moreover, \textsc{Glaser} characterizes the attacks/behaviors related to detected erasable accounts through graph analysis.
\end{abstract}

\section{Introduction}
Being the largest blockchain that supports smart contract, Ethereum has two kinds of accounts: EOA (Externally Owned Account) and contract account~\cite{li2017survey}. As a permissionless blockchain system, Ethereum allows any user to create many EOAs through their private keys. Deploying a smart contract to Ethereum will produce a contract account that contains the contract's runtime bytecodes. Every node must synchronize blockchain data, which includes blocks and StateDB (State DataBase)~\cite{8}. The StateDB stores all the accounts' state information, such as ETH balance, transaction number, runtime bytecodes, and so on~\cite{8}. 

\begin{figure}[ht]
	\centering
	\vspace*{-4ex}
	\includegraphics[width=4.50in]{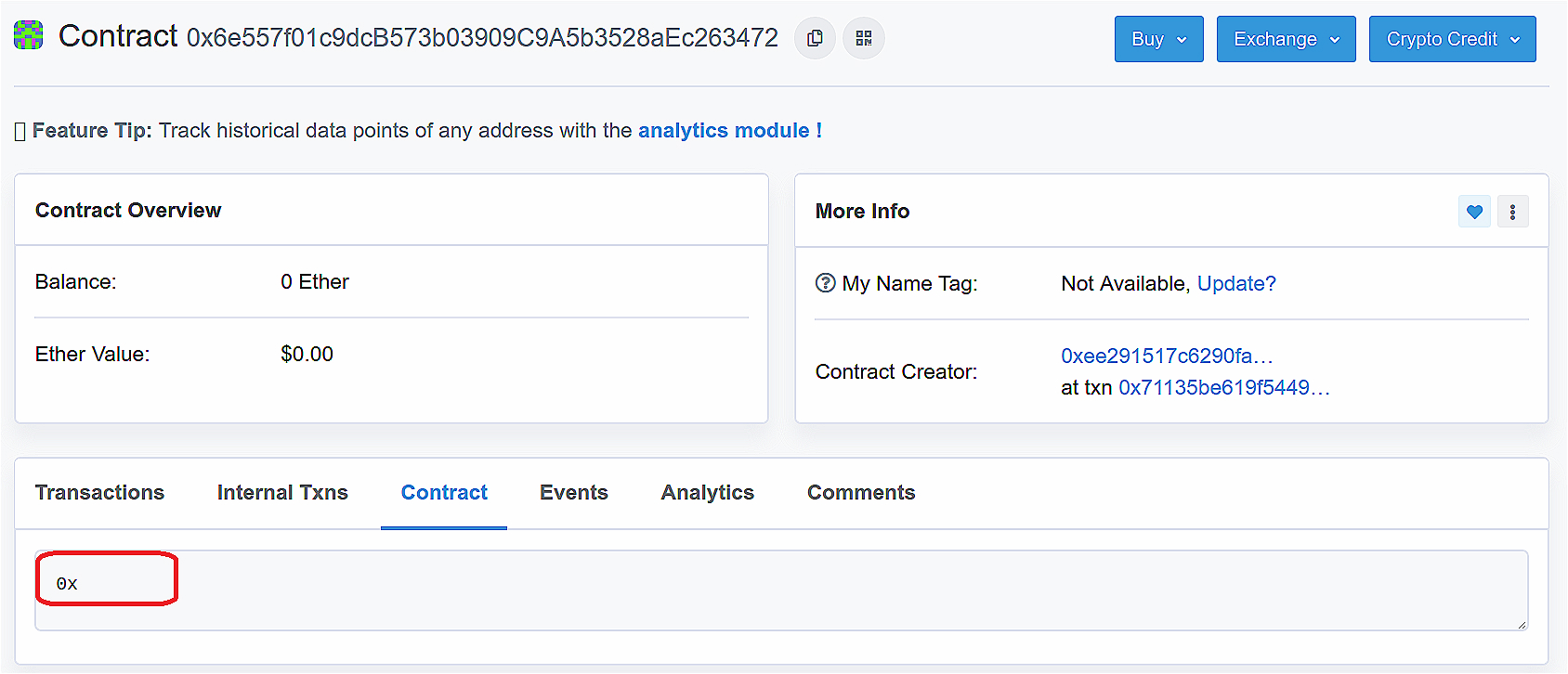}
	\vspace*{-2ex}
	\caption{One empty account detected by \textsc{Glaser}.}
	\vspace{-6ex}
	\label{example_empty}
\end{figure}

However, not all accounts should be kept. In particular, we identify three kinds of erasable contract accounts that are produced due to contracts' programming errors or attacks, and two kinds of erasable EOAs that are produced due to contracts' deployment failure or DoS (Denial of Service) attacks. Such erasable accounts not only waste system resources and affect the efficiency of blockchain, but also easily waste users' money (i.e., ETH or gas). For example, one empty account (Address: \href{https://etherscan.io/address/0x6e557f01c9dcb573b03909c9a5b3528aec263472}{\seqsplit{0x6e55..}}) discovered in this paper was created due to contract deployment failure. It wasted user's 137,552 gas when it was called because the contract's runtime bytecodes were not stored in this account, whose information is shown in Figure~\ref{example_empty}. 
We regard the worthless accounts that deserve to be removed without affecting the normal operations of users and other accounts as erasable accounts. 

Unfortunately, there lacks a systematic study on the erasable accounts that can be removed. Although some studies ~\cite{chen2020traveling}\cite{chen2018understanding} use call graph analysis to measure the control flow between contracts, their purposes are different from ours. Our work focuses on the erasable accounts that exist in Ethereum, and some of our analyzed accounts (e.g., DoS contracts) are related to interaction between contracts. There also exist some other research analyzing different kinds of security issues for smart contracts~\cite{di2019collateral} or Ethereum architecture~\cite{kiffer2018analyzing}. These research mainly focus on security issues on the contract-level and system-level of Ethereum, whose contents and purposes are different from ours.

To fill the gap, we design and implement a novel tool named \textsc{Glaser} (detectinG erasabLe AccountS in EtheReum) to discover erasable accounts by analyzing the StateDB of Ethereum. It is worth noting that marking an account as erasable just according to its liveness and balance value is improper, because an account might contain useful runtime bytecodes or its private key is owned by external user so that it cannot be removed even if it has not been used for a long time and stores no ETH. Instead, \textsc{Glaser} analyzes accounts' contents and states stored in Ethereum StateDB. In detail, it leverages program analysis techniques to discover contract accounts with worthless runtime bytecodes, and employs state field and transaction analysis to discover EOAs that no one owns their private keys. The accounts discovered by \textsc{Glaser} are worthless and deserve to be removed without affecting the normal operations of other accounts/users.

Applying \textsc{Glaser} to all Ethereum accounts, we discovered 508,482 erasable accounts, and more than 99.9\% of them are still stored in Ethereum. These erasable accounts have wasted users more than 106 million dollars and can be removed through executing \texttt{SELFDESTRUCT} operation in their runtime bytecodes by users, or removed forcibly by Ethereum officials. For example, one erasable contract account (Address: \href{https://etherscan.io/address/0xa30BCeA7E5806aC5D37D221D2F8A40642B0Bb1a6}{\seqsplit{0xa30B..}}) can be removed through transaction sent by any user, and some empty account created due to DoS attacks were already removed forcibly through hard fork by Ethereum officials~\cite{3}. This paper mainly focuses on erasable accounts' detection to help users identify erasable accounts and remind users not to call them to save money, and erasable accounts' characterization to interpret their behaviors/attacks and creation reasons. 

The main contributions of this paper are as follows:

(1) To the best of our knowledge, we conduct the \textsl{first} systematic investigation on erasable accounts in Ethereum. We propose and define five kinds of erasable accounts, i.e., three kinds of erasable contracts and two kinds of erasable EOAs.

(2) We design a novel approach to analyze the Ethereum StateDB, and implement the idea in a tool called \textsc{Glaser}, which can discover and characterize erasable contract accounts and erasable EOAs. For contract accounts, leveraging static analysis and symbolic execution, \textsc{Glaser} analyzes runtime bytecodes of contracts to detect three kinds of erasable contract accounts. For EOAs, \textsc{Glaser} analyzes their state-related attribute fields and historical transactions to discover two kinds of erasable EOAs. \textsc{Glaser} also characterizes erasable accounts through call graph and creation graph analysis.

(3) We conduct experiments to evaluate and characterize the detected erasable accounts. We analyze the 508,482 detected erasable accounts' creation time distributions. More than 99.9\% of them are still stored in Ethereum, and their transactions wasted users more than 106 million dollars. \textsc{Glaser} can remind users not to call erasable accounts and help stop further economic loss of users caused by them. Furthermore, the graph analysis of erasable accounts interprets their creation reasons, i.e., attacks, programming errors, or deployment failure.


\section{Background}
\label{sec:background}

\subsection{Ethereum}
Supporting smart contracts, Ethereum records not only  transactions but also state transitions that occur in blockchain. Ethereum contains two types of accounts, i.e., EOA and contract account~\cite{li2017survey}, which are all indexed by 20 bytes length of addresses. 

\textbf{Account's creation and usage:} Ethereum is a permissionless blockchain system, and users can create their own EOA and store ETH (native cryptocurrency in Ethereum). Users can initiate transactions by the private key corresponding to the EOA address, including ETH transfers and contract calls. 
The contract accounts are created by EOAs or other contract accounts. In addition to storing ETH, the contract account also holds the runtime bytecodes of smart contract. There are two types of bytecodes in Ethereum: \underline{runtime bytecodes} stored in contract account, and \underline{deployment bytecodes} used for contract runtime bytecodes' deployment. The contract account is not controlled by the user's private key, but by the contract's runtime bytecodes' logics.

\textbf{Account's removal:} Users can only remove contract account through executing \texttt{SELFDESTRUCT} in its runtime bytecodes. All EOAs and contract accounts without \texttt{SELFDESTRUCT} in runtime bytecodes cannot be removed by users. In addition, all erasable accounts can be removed forcibly by Ethereum officials. Although some discovered erasable accounts in this paper cannot be removed by users, our results can remind users not to call them to save money.

\textbf{StateDB:} The StateDB stores the world state of Ethereum based on accounts. For every account $a$, its state $\sigma{[a]}$ consists of four fields~\cite{8}: If $a$ is an EOA, \underline{$\sigma{[a]}_n$} stores the number of external transactions \textsl{sent from} this account. If $a$ is a contract account, $\sigma{[a]}_n$ stores the number of contracts created by this account. \underline{$\sigma{[a]}_b$} stores the balance value (in Wei) of account $a$. \underline{$\sigma{[a]}_s$} stores the root hash of Merkle tree which encodes the storage contents of the account. \underline{$\sigma{[a]}_c$} stores the runtime bytecodes of account $a$. Note that the main difference between EOA and contract account is whether its code field is empty~\cite{8}.

\subsection{Smart Contract}
In Ethereum, each node runs an EVM (Ethereum Virtual Machine), and the runtime bytecodes of contract are executed in EVM. Smart contract can be developed through several Turing complete languages, such as Solidity (the recommended language), Serpent, and Vyper~\cite{li2017survey}.

\textbf{Execution:} When a smart contract is deployed in Ethereum, users can invoke its external functions through transactions. Note that we describe transactions sent from EOAs as \underline{external transactions}, and message-calls sent from contract accounts as \underline{internal transactions} in this paper. Gas is the basic unit of resource consumption for transactions in Ethereum~\cite{chen2020}. Before users initiate transactions, they all need to pay a certain amount of gas. When the smart contract is running in EVM, each opcode corresponds to a certain amount of gas, whose value is defined in the Ethereum Yellow Paper~\cite{8}. To prevent DoS attacks, Ethereum has modified the gas value of some specific opcodes, such as \texttt{SELFDESTRUCT}'s value was modified from 0 to 5,000 in EIP-150 (Ethereum Improvement Proposal)~\cite{chen2020}. 

\textbf{Data:} The smart contracts' execution in EVM involves three forms of data, namely storage, memory, and stack~\cite{chen2017under}. The storage data is stored in StateDB of Ethereum in the form of key-value pairs, and both key's length and value's length are 256 bits~\cite{li2017survey}. Storage is persistent and will not be released as transaction execution ends. Storage data is stored and read through two opcodes, i.e., \texttt{SLOAD} and \texttt{SSTORE}. Memory is the temporarily allocated space when smart contracts are executed in EVM, which is automatically freed as the transaction execution finishes. EVM is a 1,024 depth stack-based virtual machine, and the contracts' opcodes are all executed around the stack~\cite{li2020stan}.

\section{Erasable Accounts}
\label{sec:account}
\subsection{Erasable Contract}
\label{sec:useless_contract}
The main difference between EOA and contract account is whether its code field is empty\cite{8}. Below we introduce erasable contracts with runtime bytecodes. 

\textbf{Meaningless contract:} We analyze two kinds of meaningless contract, i.e., MC-S (Meaningless Contract with \texttt{STOP}) and MC-RS (Meaningless Contract with \texttt{REVERT} or \texttt{SELFDESTRUCT}).

MC-S refers to one particular kind of meaningless contract, whose first opcode in its runtime bytecodes is \texttt{STOP}. There exist MC-S because users incorrectly use runtime bytecodes to deploy contracts, whose creation and behavior will be analyzed in Section~\ref{sec:evaluation}. When the MC-S is called, \texttt{STOP} will halt the transaction's execution immediately. Therefore, these contracts are controlled by \texttt{STOP}, which is meaningless and may waste user's gas or ETH.

\underline{MC-S Example:} One MC-S (Address: \href{https://etherscan.io/address/0x2Ab748a546760b1EC834E164DEDE2E71C4010E1d}{\seqsplit{0x2Ab7..}}) was called with input data three times, which waste users' gas. Their input data were not processed before the related transactions were halted by \texttt{STOP}. Furthermore, this meaningless contract was transferred ETH through transactions twice. Because the MC-S is controlled by \texttt{STOP}, the total of more than 0.042 ETH stored in this account can never be transferred out, which results in users' money waste. 

MC-RS refers to contract that has \texttt{REVERT} or \texttt{SELFDESTRUCT} opcode in its first basic block. A basic block means a series of sequential opcodes without any control flow operation (e.g., \texttt{JUMP}, \texttt{STOP})~\cite{chen2020}. The first basic block is the program entrance and every call to the contract will execute it. Most MC-RS are deployed by malicious contracts through internal transactions (i.e., sent from contract). However, MC-RS is meaningless because any call to MC-RS will invoke \texttt{REVERT} or \texttt{SELFDESTRUCT}. \texttt{REVERT} ends runtime bytecodes' execution and reverts state changes of the call. \texttt{SELFDESTRUCT} removes the contract account from blockchain.

\begin{figure}[ht]
	\centering
	\vspace*{-5ex}
	\includegraphics[width=3.40in]{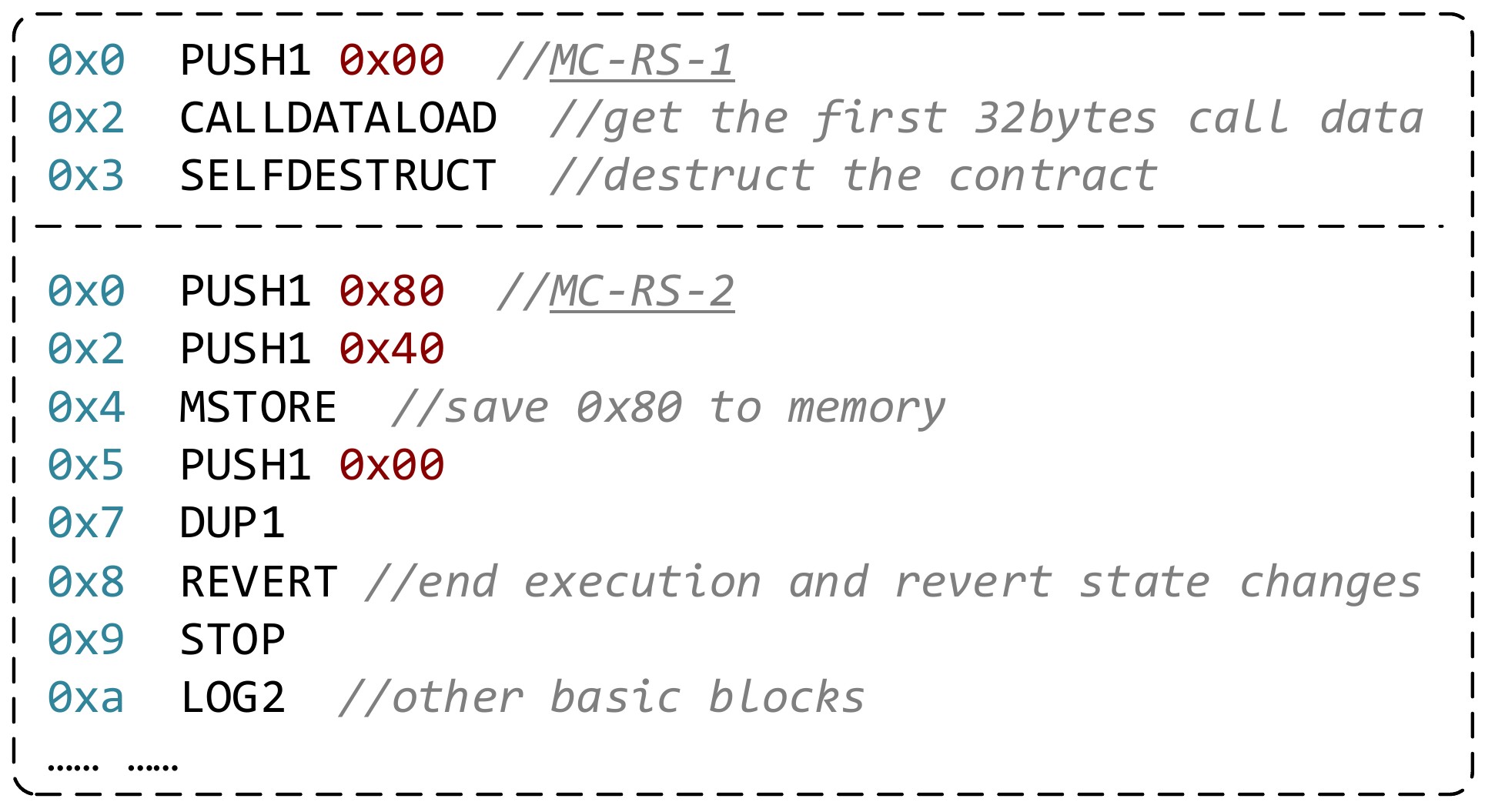}
	\vspace*{-2ex}
	\caption{Snippets of two MC-RS (i.e., with \texttt{REVERT} or \texttt{SELFDESTRUCT} in first basic blocks).}
	\vspace{-5ex}
	\label{meaningless_contract}
\end{figure}
\underline{MC-RS Example:} The snippets of two MC-RS are shown in Figure~\ref{meaningless_contract}. There are only three operations in the first MC-RS (Address: \href{https://etherscan.io/address/0xa30BCeA7E5806aC5D37D221D2F8A40642B0Bb1a6}{\seqsplit{0xa30B..}}). This contract can be exploited by attacker to steal ETH through setting his own EOA address in the call data. However, this contract is meaningless. Because any call to it will invoke \texttt{SELFDESTRUCT} and transfer out the ETH stored in it. The second MC-RS (Address: \href{https://etherscan.io/address/0x7770A80851A266e717dC93A194A7eC0875214293}{\seqsplit{0x7770..}}) will invoke \texttt{REVERT} during any call to it. Furthermore, the operations after its first basic block will never execute. Any call to the contract will execute operations from \texttt{0x0} to \texttt{0x8}, which is meaningless and waste gas.

\textbf{Stack or opcode error contract:} EVM is a virtual machine based on 1,024 depth stack, and the stack will definitely overflow (push more than 1,024 items into stack) or underflow (pop item from empty stack) if the stack error contract is called. Before EIP-150 increased the gas cost of \texttt{CALL} from 40 to 700, the attacker may exploit stack overflow through recursive call depth attack~\cite{li2017survey}. Nowadays, although stack overflow is hard to occur, stack underflow still exists due to program writing errors.

\begin{figure}[ht]
	\centering
	\vspace*{-5ex}
	\includegraphics[width=2.80in]{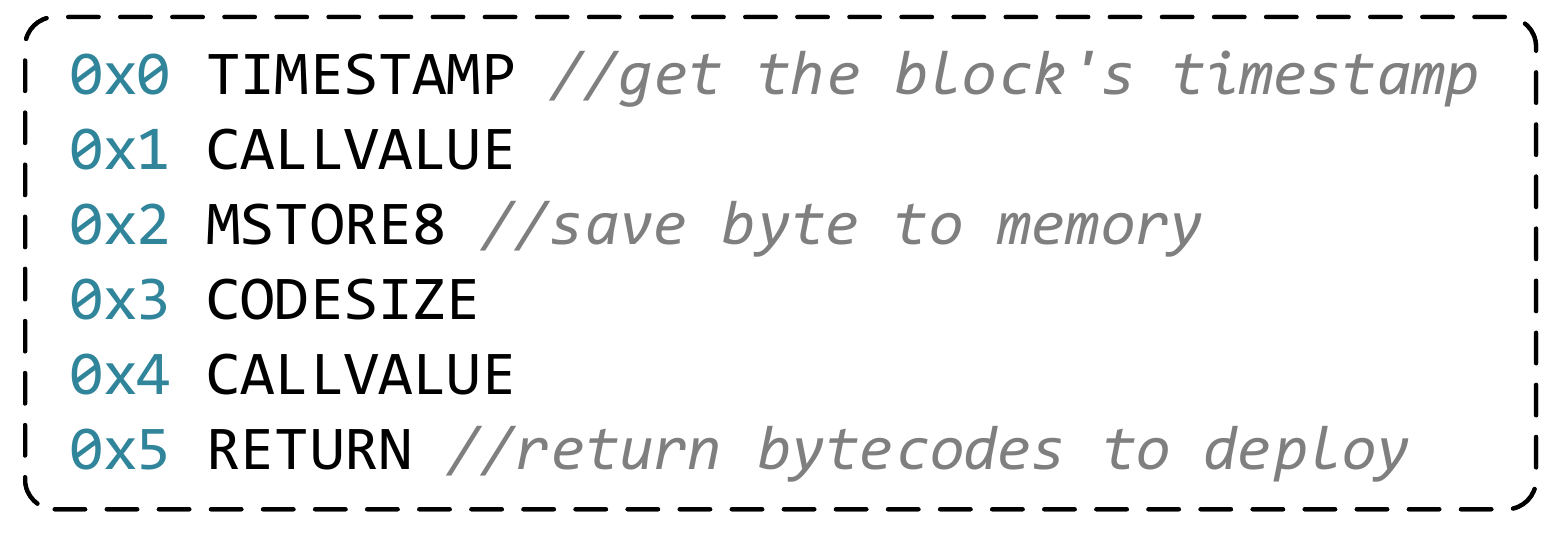}
	\vspace*{-3ex}
	\caption{The deployment bytecodes of one stack error contract.}
	\vspace{-6ex}
	\label{stack_error_contract}
\end{figure}

\underline{Stack Error Contract Example:} One transaction (Hash: \href{https://etherscan.io/tx/0x9518bcde68b522a4521c3eeade8fa461af16b5c7f0d1529d7ead27663d4e5092}{\seqsplit{0x9518..}}) encountered ``Stack Underflow'' error and exhausted its gas, due to its contract deployment-related codes. Moreover, runtime bytecodes' contents may be related to some uncontrollable factors, which may also produce stack error contracts. One example of stack error contract's deployment bytecodes is shown in Figure~\ref{stack_error_contract} (Related transaction: \href{https://etherscan.io/tx/0xf7db99fb4524133839915b8e08914dae0f9bcecd6847691e4dd2ce8ead61e420}{\seqsplit{0xf7db..}}). In program counter \texttt{0x5}, it returns runtime bytecodes to deploy, whose first byte is related to the current block's timestamp (in program counter 0x0 to 0x1). At last, one stack error contract (Address: \href{https://etherscan.io/address/0x7a0352aa3231d2255a96113b619057994341069e}{\seqsplit{0x7A03..}}) was deployed, and its first operation in runtime bytecodes is \texttt{DIV}, which will result in stack underflow. 



Developers can use high-level languages or directly write bytecodes to develop smart contract. However, due to programming errors, some runtime bytecodes deployed in blockchain cannot be disassembled to correct opcodes. If there exist unknown opcodes that cannot be recognized by EVM, it will encounter ``Bad Instruction Error''. Opcode error contract refers to contract that has unknown opcode in the first basic block, which will encounter error during any call to it.

\underline{Opcode Error Contract Example:} The first two bytes of one contract's runtime bytecodes are \texttt{0xd929}, which cannot be correctly disassembled to opcodes of EVM. Because the unknown opcodes exist in its first basic block, all transactions calling to it encountered ``Bad Instruction Error'' (Address: \href{https://etherscan.io/address/0x526634cde83e541ba851a402e5c85bd0838505eb/advanced#internaltx}{\seqsplit{0x5266..}}). The transaction with ``Bad Instruction Error'' exhausts gas, halts execution and reverts state changes~\cite{8}.

\textbf{DoS contract:} We analyze two kinds of DoS related contracts: attacked Parity wallets, and malicious contracts exploited for DoS attacks. If contract \texttt{A} hardcodes and calls contract \texttt{B}'s address to execute, and \texttt{B} is removed, \texttt{A} will be a dependency error contract without normal service. In November of 2017, the attacker escalated his privilege and removed Parity's multi-sig library contract (Address: \href{https://etherscan.io/address/0x863df6bfa4469f3ead0be8f9f2aae51c91a907b4}{\seqsplit{0x863d..}}), which caused all Parity wallets that depend on it out of service. Note that calling to a removed contract will just return 1 (means no error or exception), and users cannot verify if it is out of service through return value. If users knew in advance that their wallets were out of service, they would not use them anymore to deduce financial losses. Etherscan only marks part of attacked Parity wallets, we attempt to detect more of them.


In 2016, the attacker exploited malicious contracts to initiate DoS attack for Ethereum~\cite{li2017survey}. The attacker executes massive particular operations (e.g., \texttt{EXTCODESIZE}, \texttt{DELEGATECALL}), which consume low gas but high system resources. The DoS attack leads to low nodes' data synchronization and transaction execution. The Ethereum official modified many operations' gas values in EIP-150~\cite{li2017survey} to repair related vulnerabilities. 

\begin{figure}[ht]
	\centering
	\vspace*{-5ex}
	\includegraphics[width=3.40in]{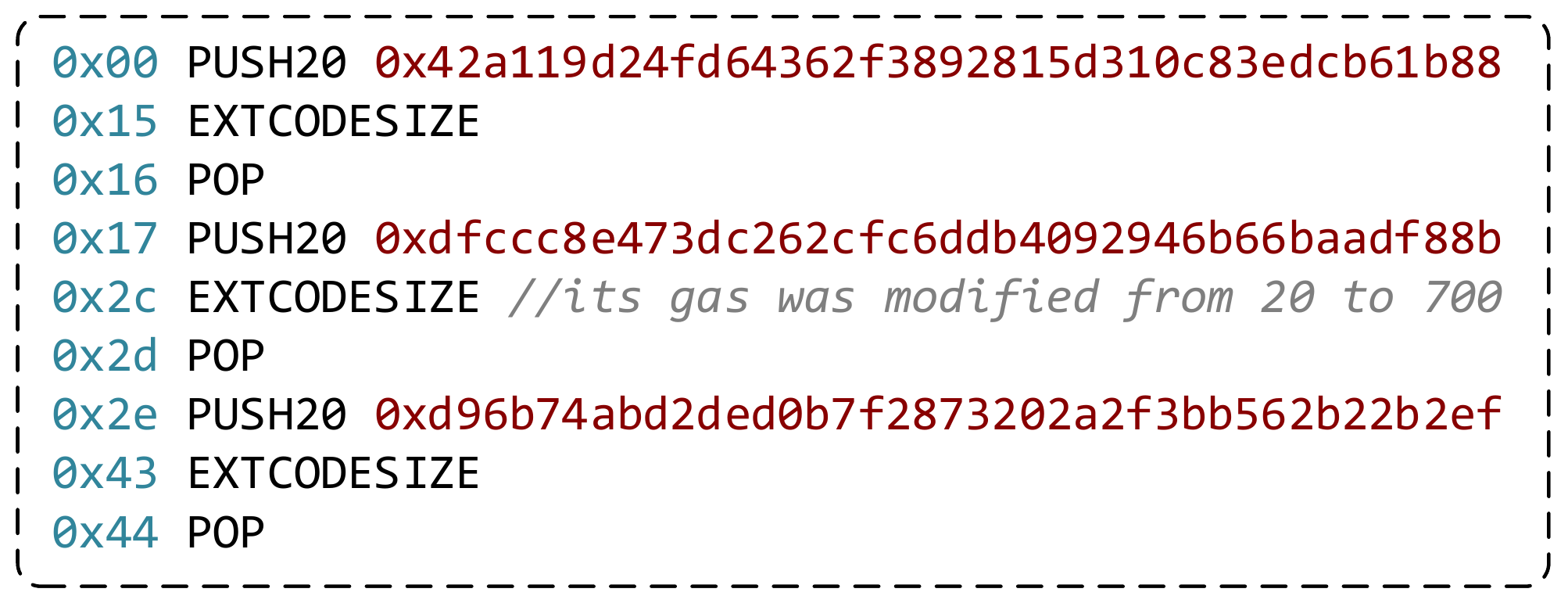}
	\vspace*{-3ex}
	\caption{Snippets of one malicious DoS contract.}
	\vspace{-9ex}
	\label{dos_contract}
\end{figure}

\underline{Malicious DoS Contract Example:} We analyze the malicious DoS contracts and discover that they have similar patterns. These malicious contracts just have one basic block in their runtime bytecodes. In the basic block, there are many particular operations that consume low gas but high system resources. For example, one malicious DoS contract's snippets are shown in Figure~\ref{dos_contract} (Address: \href{https://etherscan.io/address/0x792218d8bbe00fb81296236b014Fb14af2DA385B}{\seqsplit{0x7922..}}) with 200 \texttt{EXTCODESIZE} in the only basic block of the contract. 

\subsection{Erasable EOA}
\label{sec:erasable-eoa}

\textbf{Empty account:} The empty account has the following features: \ding{182}\uline{zero value balance}, \ding{183}\uline{zero value nonce}, and \ding{184}\uline{empty code}. Whether the account has code is the main difference between contract account and EOA, and we classify empty accounts into erasable EOA. Ethereum officials have only cleaned up the empty accounts created during the DoS attack exploiting \texttt{SELFDESTRUCT}~\cite{3}. However, there still exist empty accounts due to  contract deployment failure. Before EIP-2, it will create an empty account if the contract deployment transaction does not succeed (e.g., out of gas). After EIP-2, it will fail with error and do not create empty accounts anymore. The creation process of empty accounts denotes that they are not controlled by runtime bytecodes or external users, which results in their uselessness. The empty account may result in gas waste, because users may incorrectly think runtime bytecodes are deployed in these accounts.

\underline{Example:} One empty account (Address: \href{https://etherscan.io/address/0x6e557f01c9dcB573b03909C9A5b3528aEc263472}{\seqsplit{0x6e55..}}) has been called many times, which wastes users' gas. We analyze all the input data of the related transactions, whose first four bytes are all function signatures. That is to say, all these transactions were intended to invoke the functions in runtime bytecodes.

\textbf{DoS EOA:} The DoS EOA has the following features: \ding{182}\uline{1 Wei value balance}, \ding{183}\uline{zero value nonce}, \ding{184}\uline{empty code}, \ding{185}\uline{zero historical external transaction}, and \ding{186}\uline{one historical internal transaction without error}. The differences between empty account and DoS EOA are their balance value and creation process. DoS EOAs are created through internal transactions sent from contracts. Massive DoS EOAs were created during the DoS attack in 2016, whose creation will be analyzed in Section~\ref{sec:eval_qua}. The attacker created DoS EOAs through smallest financial cost (i.e., 1 Wei), and all of these accounts' addresses were generated through computation in runtime bytecodes, whose process denotes their uselessness (detail in Section~\ref{sec:graph}). The existence of massive DoS EOAs increases the StateDB size, resulting in the waste of disk resources and nodes' difficulty in syncing data.

\underline{Example:} One transaction (Hash: \href{https://etherscan.io/tx/0x1aa87a25df792f1dacacbc194e3963a0cbcf950ede1d60e679500b40d9589b17}{\seqsplit{0x1aa8..}}) detected by \textsc{Glaser} created ten DoS EOAs through internal transactions. Note that 1 Wei (1 ETH = $10^{18}$ Wei) is the smallest cryptocurrency unit in Ethereum, which cannot even buy 1 gas. The recommended gas price is 61 GWei~\cite{14} (1 GWei = $10^9$ Wei), which can be set in transaction by users.

\section{GLASER}
\label{sec:glaser}
To analyze the StateDB, we synchronize the blockchain with ``fat-db=on'' option through Parity client, which can build appropriate information to allow enumeration of all accounts. Then we export the StateDB as plain text file through Parity and leverage \textsc{Glaser} to traverse StateDB data to detect erasable accounts. The overview of \textsc{Glaser}'s architecture is shown in Figure~\ref{architecture}, which mainly consists of three modules:
\begin{figure}[ht]
	\centering
	\vspace*{-6ex}
	\includegraphics[width=4.50in]{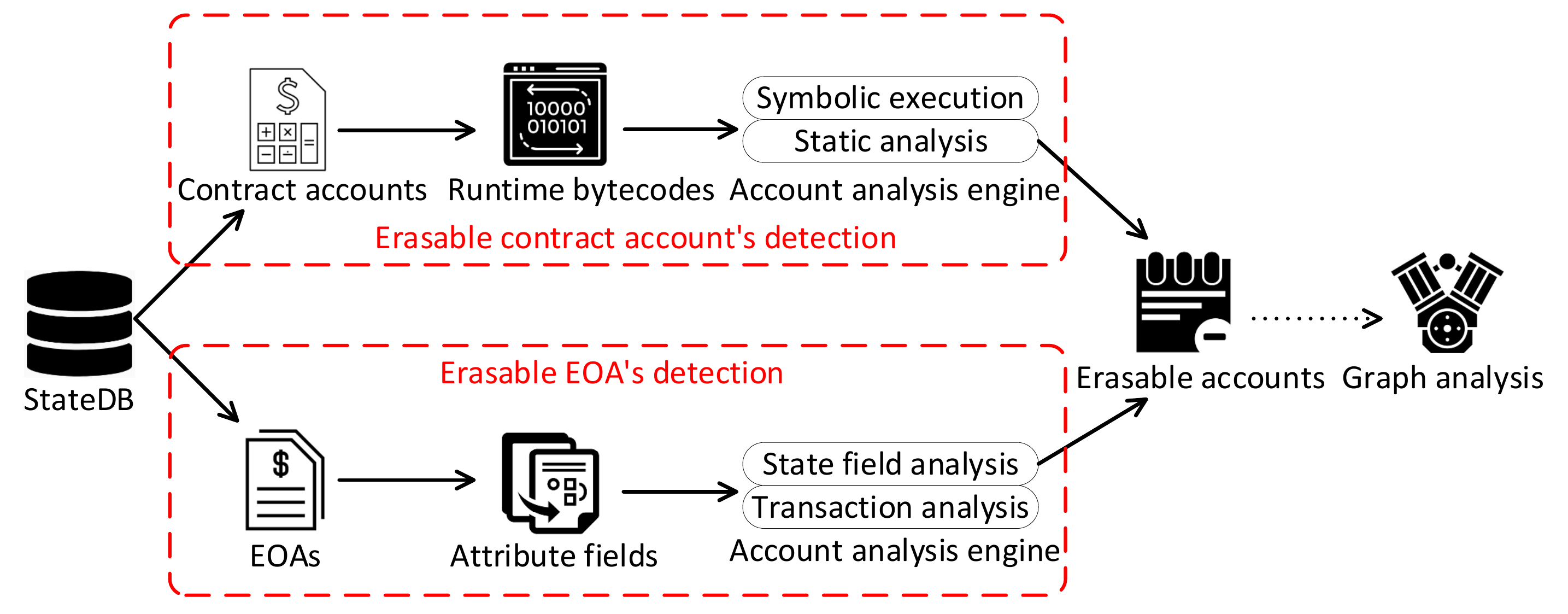}
	\vspace*{-3ex}
	\caption{Overview of \textsc{Glaser}'s architecture.}
	\vspace{-6ex}
	\label{architecture}
\end{figure}

(1) Erasable contract account detection. In this module, \textsc{Glaser} detects three kinds of erasable contract accounts: meaningless contracts, stack/opcode error contracts, and DoS contracts. According to their respective characteristics, we leverage different techniques, which mainly include runtime bytecodes' static analysis and symbolic execution.

(2) Erasable EOA detection. In this module, \textsc{Glaser} detects two kinds of erasable EOAs: empty accounts, which are produced due to contract deployment failure; and DoS EOAs, which are produced due to DoS attacks. We mainly leverage state field and transaction analysis to discover erasable EOAs.

(3) Graph analysis for erasable accounts. For the detected erasable accounts, \textsc{Glaser} characterizes their behaviors/attacks through call graph analysis and creation graph analysis, whose details will be described in Section~\ref{sec:graph}.

\subsection{Erasable Contract Detection}
\textbf{Meaningless contract:} \textsc{Glaser} leverages runtime bytecodes' static analysis to detect two kinds of meaningless contract, i.e., MC-S and MC-RS. Static analysis refers to techniques that examine codes without attempting to execute them~\cite{chess2004static}. \textsc{Glaser} statically analyzes contracts' runtime bytecodes to detect MC-S. In detail, it intercepts runtime bytecodes' first byte to judge whether it is \texttt{0x00}, which is the hex code for \texttt{STOP}. If one contract starts with \texttt{0x00} byte in its runtime bytecodes, it will be tagged as MC-S. \textsc{Glaser} also statically analyzes contracts' runtime bytecodes to detect MC-RS. First, it disassembles contract's runtime bytecodes to acquire the opcodes. Second, it splits the opcodes into different basic blocks, which end with specific control flow operations (i.e., \texttt{STOP}, \texttt{JUMP}, \texttt{JUMPI}, \texttt{RETURN}, \texttt{SELFDESTRUCT}, \texttt{REVERT}). Third, it analyzes the first basic block. If \texttt{REVERT} or \texttt{SELFDESTRUCT} exists in the first basic block, it will tag the contract as MC-RS.

\textbf{Stack or opcode error contract:} \textsc{Glaser} leverages symbolic execution and runtime bytecodes' static analysis to detect stack/opcode error contracts. Symbolic execution uses symbolic values as inputs to simulate the process of program execution~\cite{luu2016making}. The detection process of stack error contract is divided into three steps. First, \textsc{Glaser} acquires the opcodes of contract's runtime bytecodes. Second, it splits the opcodes into different basic blocks and extracts the runtime bytecodes corresponding to the first basic block. Third, it symbolically executes the extracted runtime bytecodes leveraging \textsc{Oyente}~\cite{luu2016making}, which is a symbolic execution engine. If the symbolic execution process encounters ``Stack Underflow'', it will tag the contract as stack error contract. For opcode error contract, \textsc{Glaser} disassembles contract's runtime bytecodes into opcodes and split them into basic blocks. Then \textsc{Glaser} detects whether there exist unknown opcodes in its first basic block. If unknown opcode exists in its first basic block, the contract will be tagged as opcode error contract.


\textbf{DoS contract:} \textsc{Glaser} leverages symbolic execution and runtime bytecodes' static analysis to detect DoS contracts. \textsc{Glaser} detects attacked Parity wallet contracts leveraging symbolic execution techniques. \textsc{Glaser} analyzes four related operations for contracts' interaction, i.e., \texttt{CALL}, \texttt{CALLCODE}, \texttt{DELEGATECALL}, \texttt{STATICCALL}. If the symbolic execution encounters anyone of these operations, it extracts the second item of the stack $\mu_{s}[1]$, which is used as the address of contract being called. If $\mu_{s}[1]$ is a real value that matches the address of removed Parity multi-sig library, it will tag this contract as attacked Parity wallet. For malicious DoS contract, \textsc{Glaser} disassembles contract's runtime bytecodes into opcodes and split them into basic blocks. Then \textsc{Glaser} analyzes the number and content of basic block. If one contract has only one basic block and has more than 100 DoS related operations, \textsc{Glaser} will tag it as malicious DoS contract. We analyze seven DoS related operations: \texttt{EXTCODESIZE}, \texttt{EXTCODECOPY}, \texttt{BALANCE}, \texttt{CALL}, \texttt{DELEGATECALL}, \texttt{CALLCODE}, \texttt{SELFDESTRUCT}.

\subsection{Erasable EOA Detection}
\textbf{Empty account:} \textsc{Glaser} leverages account state field analysis and transaction analysis to detect empty accounts. The detection process of empty accounts is divided into two steps. First, \textsc{Glaser} analyzes the account attribute fields to detect possible empty accounts, which should satisfy the three features described in Section~\ref{sec:erasable-eoa}. Second, \textsc{Glaser} analyzes the historical transaction of the detected empty accounts in the first step, to verify that they are created due to contract deployment failure. In detail, it analyzes the oldest transaction related to the accounts detected in the first step. If one account's oldest transaction is used for contract deployment, \textsc{Glaser} will tag it as erasable empty account. 

\textbf{DoS EOA:} \textsc{Glaser} leverages account state field analysis and transaction analysis to detect DoS EOA. Similar to the detection of empty accounts, detection process of DoS EOAs is divided into two steps. First, \textsc{Glaser} analyzes the account attribute fields to detect possible DoS EOAs, which should satisfy the first three features described in Section~\ref{sec:erasable-eoa}. Second, \textsc{Glaser} analyzes the historical transaction of the detected DoS EOAs in the first step, to verify that they are created through internal transactions sent from contracts. In detail, we set relatively strict conditions to verify DoS EOAs in this step. \textsc{Glaser} analyzes their historical external transactions and internal transactions. If one account has no external transaction and only one internal transaction without error (i.e., sent 1 Wei to create this account), we can conclude that it is an erasable DoS EOA. There might exist massive internal transactions with ``Out of Gas Error'', which were used for DoS attacks.

\section{Evaluation}
\label{sec:evaluation}
We carry out experiments to answer the following research questions: \emph{RQ1 Quantity:} How many each kind of erasable accounts can be detected through \textsc{Glaser}? \emph{RQ2 Accuracy:} To what extent can \textsc{Glaser} accurately detect erasable accounts? \emph{RQ3 Waste:} How much money lost due to erasable accounts?

\subsection{RQ1 Quantity}
\label{sec:eval_qua}
In this section, we evaluate the quantity statistics of erasable accounts detected through \textsc{Glaser}. Furthermore, we analyze the creation time distribution of the detected erasable accounts.

\begin{table}[ht!]
	\centering
	\vspace*{-4ex}
	\scriptsize
	\caption{Quantity statistics of erasable accounts detected through \textsc{Glaser}.}
	\vspace{-2ex}
	\label{tab_quantity}
	\begin{tabular}{|c|c|c|c|c|}
		\hline
		\textbf{Cat.}&\textbf{Taxonomy}&\textbf{Quantity}&\textbf{Erasable accounts}&\textbf{Quantity}\\ \hline
		\multirow{3}*{\ding{182}}&\multirow{3}*{Erasable contract}&\multirow{3}*{481,087}&Meaningless contract&479,153 \\ \cline{4-5}
		~&~&~&Stack/opcode error contract&150 \\ \cline{4-5}
		~&~&~&DoS contract&1,784 \\ \hline
		\multirow{2}*{\ding{183}}&\multirow{2}*{Erasable EOA}&\multirow{2}*{27,395}&Empty account&195 \\ \cline{4-5}
		~&~&~&DoS EOA&27,200 \\ \hline
	\end{tabular}
	\vspace{-6ex}
\end{table}
We have exported the StateDB of Ethereum and detect erasable accounts leveraging \textsc{Glaser}, whose quantity statistics are shown in Table~\ref{tab_quantity}. We discover 481,087 erasable contracts and 27,395 erasable EOAs respectively. All the five specific kinds of detected erasable accounts' addresses are published on\textcolor{blue}{\url{https://figshare.com/articles/dataset/11516694}}. For the 1,784 DoS contracts, we detect 658 different contracts hardcode and call the removed Parity multi-sig library, while Etherscan only tags 153 of them. Because most users leverage high-level languages to develop contracts, there exists a small quantity of stack/opcode error contracts. Because Ethereum officials have already repaired the bug of empty account's creation due to contract deployment failure, the discovered empty accounts' quantity is small.

To measure the number of erasable accounts at different time, we analyze their historical transactions to acquire their creation time. The analysis of accounts' creation time is divided into two steps. First, we crawl all the historical transactions related to the detected erasable accounts through Geth RPC APIs. Second, we filter out the oldest transaction of each account and acquire the timestamp of this transaction, which is the creation time of this account. 

\begin{figure}[ht]
	\centering
	\vspace*{-5ex}
	\includegraphics[width=4.10in]{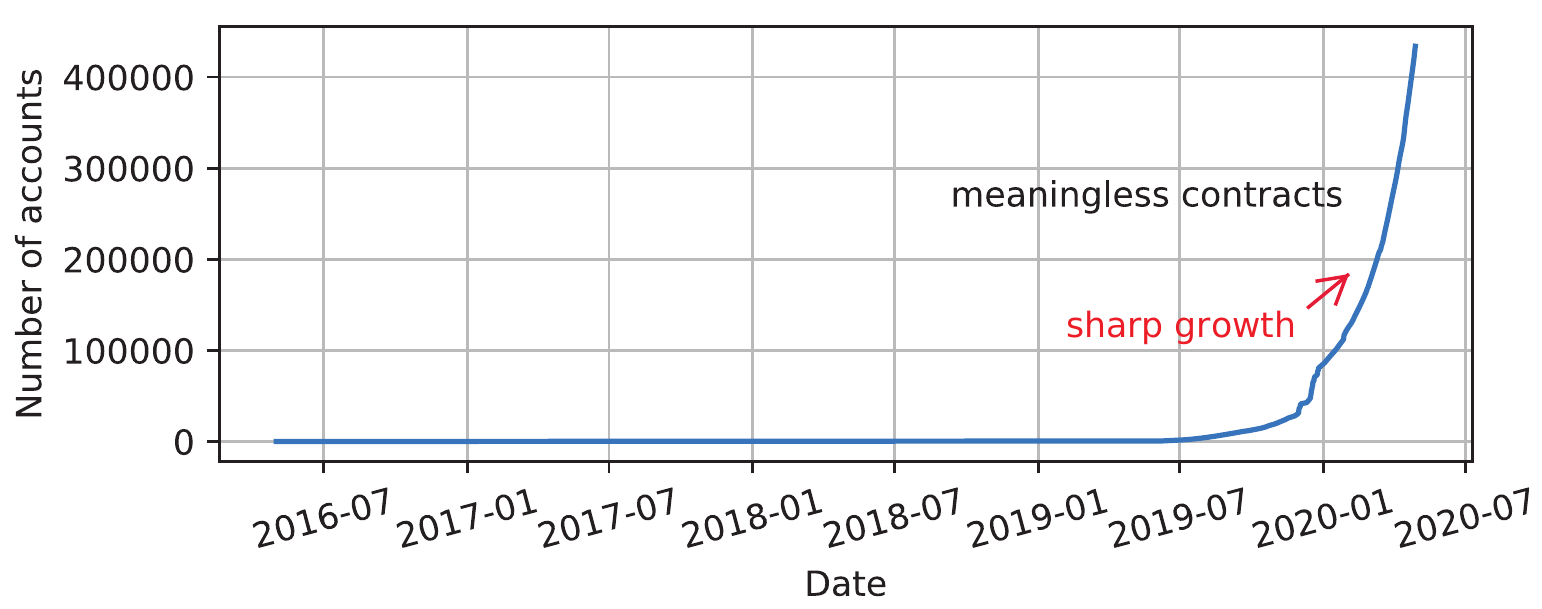}
	\vspace*{-3ex}
	\caption{Cumulative quantity distribution of meaningless contracts.}
	\vspace{-6ex}
	\label{time_meaningless}
\end{figure}
\begin{figure}[ht]
	\centering
	\vspace*{-5ex}
	\includegraphics[width=4.00in]{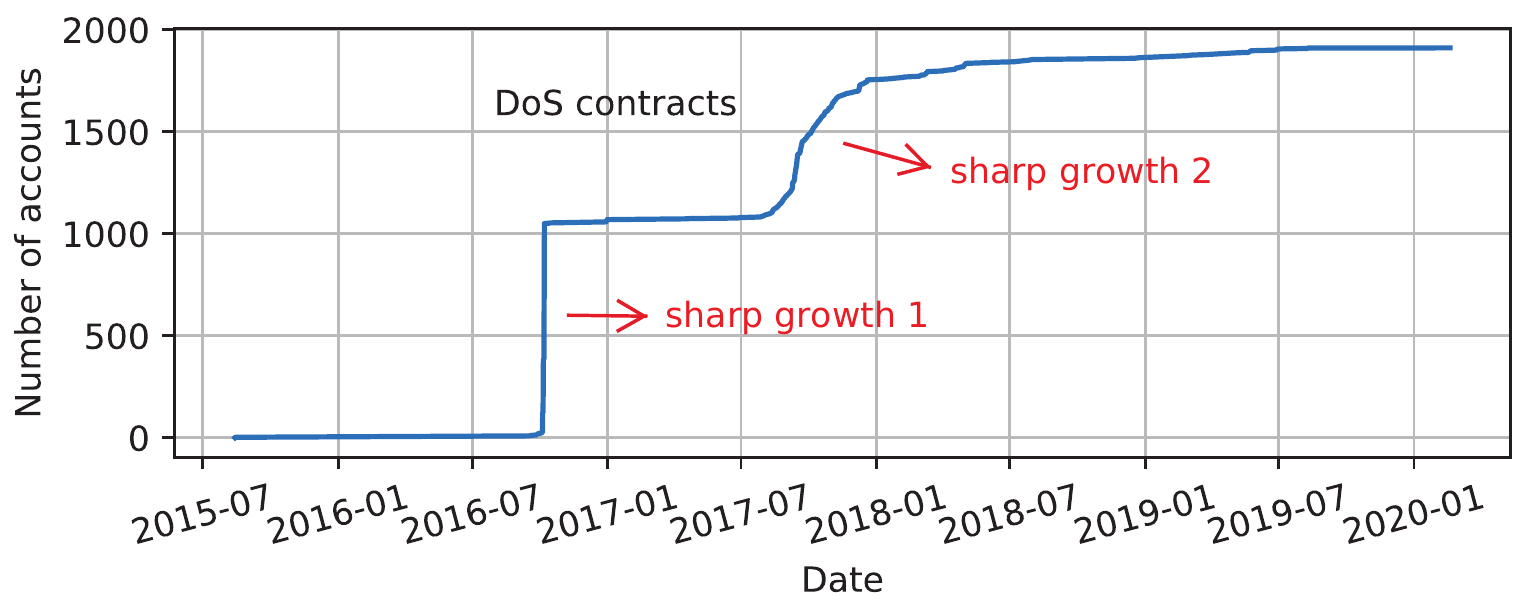}
	\vspace*{-3ex}
	\caption{Cumulative quantity distribution of DoS contracts.}
	\vspace{-5ex}
	\label{time_dos}
\end{figure}
The cumulative quantity distribution of meaningless contracts at different time is shown in Figure~\ref{time_meaningless}. Before July of 2019, the quantity of meaningless contracts is small. Because most meaningless contracts are MC-S and they are directly created by users through EOA. For example, one user (Address: \href{https://etherscan.io/address/0x3ff51120D34f4318B6aff85DbCa5481DbF03f40B}{\seqsplit{0x3ff5..}}) created 9 MC-S with totally same runtime bytecodes around February of 2018. When the user realized his irrational behavior, he did not create MC-S any more. After November of 2019, some active malicious contracts are massively called, which leads to the quantity sharp growth of meaningless contracts (i.e., MC-RS). For example, one Ponzi contract (Address: \href{https://etherscan.io/address/0x7C20218efC2e07C8Fe2532fF860D4A5d8287cB31}{\seqsplit{0x7C20..}}) created many MC-RS before April of 2020 through internal transactions (i.e., sent from contract account). Because most users leverage high-level languages to develop contracts, there exists a small quantity of stack/opcode error contracts. Their deployment time distribution does not have clear trends or characteristics.


The cumulative quantity distribution of DoS contracts at different time is shown in Figure~\ref{time_dos}. There are two sharp growth periods for DoS contracts. The first period is around October of 2016, the attacker deployed more than 1k malicious DoS contracts and sent massive transactions to them, leading to external transactions' slow execution. The second period is around November of 2017, the Parity’s multi-sig library contract was attacked and removed during this period, which produced 658 dependency error wallets without service.

\begin{figure}[ht]
	\centering
	\vspace*{-4ex}
	\includegraphics[width=4.00in]{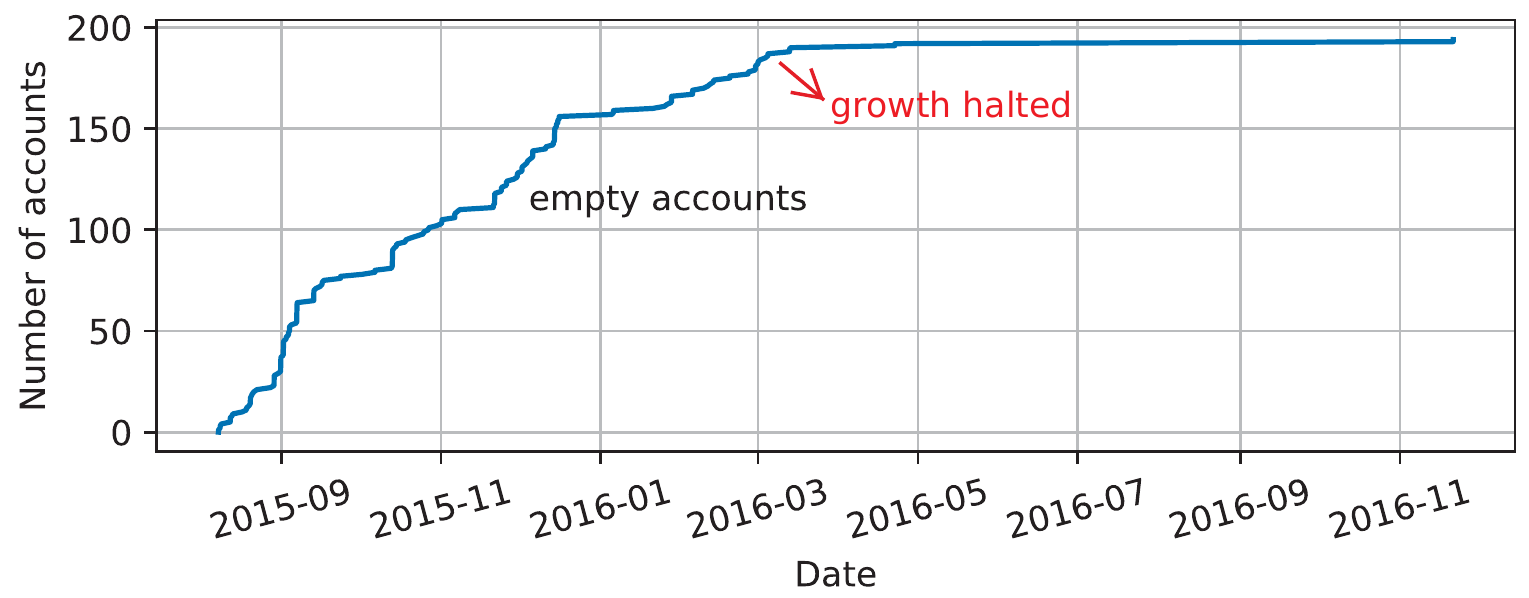}
	\vspace*{-3ex}
	\caption{Cumulative quantity distribution of empty accounts.}
	\vspace{-5ex}
	\label{time_empty}
\end{figure}

\begin{figure}[ht]
	\centering
	\vspace*{-6ex}
	\includegraphics[width=4.05in]{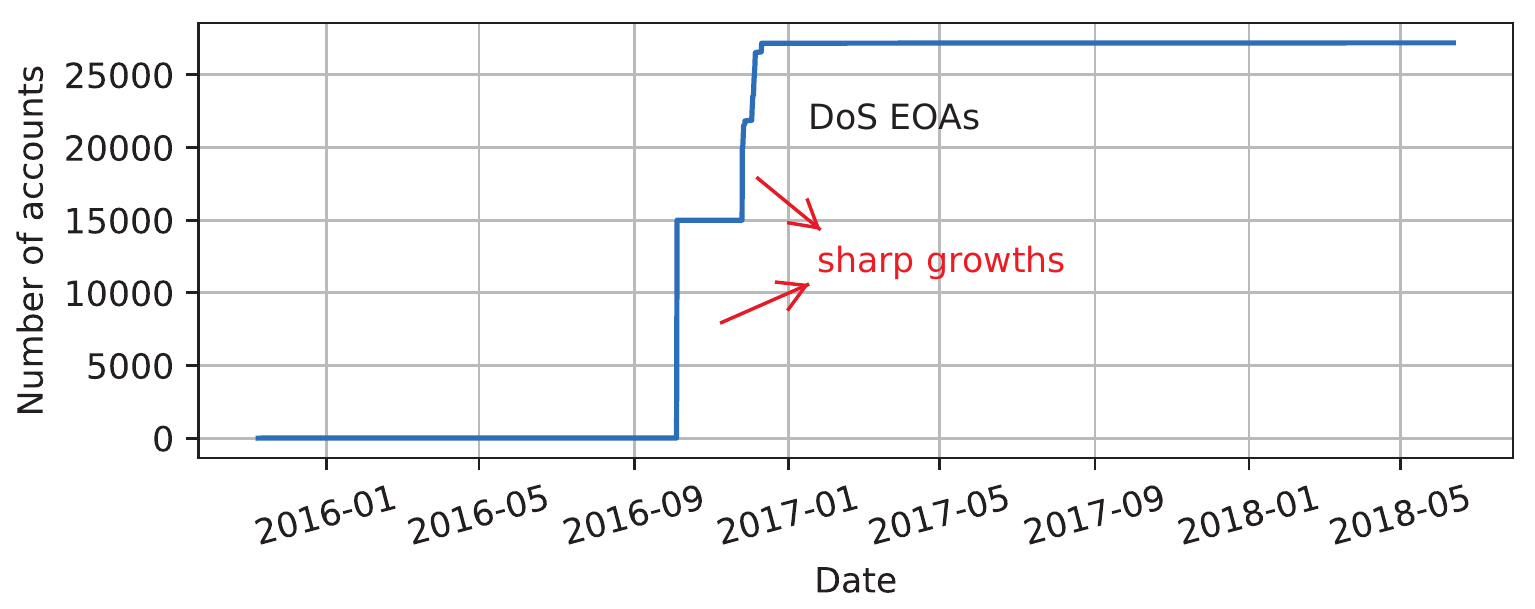}
	\vspace*{-3ex}
	\caption{Cumulative quantity distribution of DoS EOAs.}
	\vspace{-8ex}
	\label{time_1wei}
\end{figure}

The cumulative quantity distribution of empty accounts at different time is shown in Figure~\ref{time_empty}. Because the Ethereum officials have repaired the bug of empty accounts' creation due to contract deployment failure and cleaned up the empty accounts produced due to DoS attacks, the growing of their cumulative quantity is halted around March of 2016. The cumulative quantity distribution of DoS EOAs is shown in Figure~\ref{time_1wei}. There is a sharp growth period of DoS EOAs' quantity around November of 2016. According to analysis, the attacker (One exploited account: \href{https://etherscan.io/address/0xeec2a1ee6ee942596b6e255d24d38c0a9338cfef}{\seqsplit{0xeec2..}}) created massive DoS EOAs during/after the DoS attacks of empty accounts' creation exploiting \texttt{SELFDESTRUCT}~\cite{3}.

\noindent\underline{\textbf{Answer to RQ1 (Quantity):}} We have discovered 508,482 erasable accounts, whose quantity distributions at different time reflect their creation reasons.

\subsection{RQ2 Accuracy}
In this section, we evaluate the accuracy of erasable accounts detected through \textsc{Glaser}, whose statistics are shown in Table~\ref{sta_accu}. We evaluate the accuracy of erasable accounts in two primary aspects. First, we analyze whether the detected erasable accounts are still stored in Ethereum. Second, we analyze their transactions to verify their uselessness.

\textbf{Storage:} Because it is difficult to traverse accounts in its changing StateDB, we export the StateDB to offchain and execute \textsc{Glaser} on it. Therefore, there exists possibility that the detected erasable accounts are already removed or cleaned up in the newest StateDB. We leverage Etherscan, which is a real-time Ethereum block explorer, to verify their existence. We discover that 99.9\% (479,150/479,153) detected erasable contract accounts still store runtime bytecodes. 3 MC-RS contracts are removed through executing \texttt{SELFDESTRUCT}. All the 195 empty accounts can be still normally retrieved without special tagging, and all the 27,200 detected DoS EOAs still store ETH. Therefore, more than 99.9\% of the detected erasable accounts are still stored in the latest StateDB.

\begin{table*}[ht!]
	\centering
	\vspace*{-4ex}
	\scriptsize
	\caption{Statistics of erasable accounts' accuracy and waste evaluation.}
	\vspace{-2ex}
	\label{sta_accu}
	\begin{tabular}{|c|c|c|c|c|c|c|}
		\hline
		\textbf{Erasable account}&\textbf{Quantity}&\textbf{Storage}&\textbf{Ext. tr.}&\textbf{Int. tr.}&\textbf{Gas}&\textbf{ETH}\\ \hline
		\ding{182} \textsc DoS contract&1,784&100\%~\ding{51}&26,474&7,707,646&50,497,619,162&515,035.16ETH \\ \hline
		\ding{183} \textsc Meaningless contract&479,153&99.9\%~\ding{51}&2,080&490,611&36,996,614,413&274.97ETH \\ \hline
		\ding{184} \textsc Stack/opcode error cont.&150&100\%~\ding{51}&141&157,513&854,099,555&0 \\ \hline
		\ding{185} \textsc Empty account&195&100\%~\ding{51}&237&5&79,786,061&0 \\ \hline
		\ding{186} \textsc DoS EOA&27,200&100\%~\ding{51}&0&1,163,763&1,180,693,660&27,200 Wei \\ \hline
	\end{tabular}
	\vspace{-6ex}
\end{table*}

\textbf{Uselessness:} In the following, we analyze the detected erasable accounts' transactions to verify their uselessness. All the below analyzed transactions' data is published on\textcolor{blue}{\url{https://figshare.com/articles/dataset/11518017}}.

For the 1,784 DoS contracts, we crawl all of their 26,474 external transactions and 7,707,646 internal transactions. According to the timestamp of Parity multi-sig library's removal (Transaction hash: \href{https://etherscan.io/tx/0x47f7cff7a5e671884629c93b368cb18f58a993f4b19c2a53a8662e3f1482f690}{\seqsplit{0x47f7..}}), we extract 920 external transactions of attacked Parity wallets that occurred after the attack. Apart from pure ETH transfers, there are 789 external transactions calling wallets' functions. Because calling to a removed contract does not result in failure or exception, we debug these transactions for analysis. We acquire these transactions' execution traces through Geth API \texttt{debug\_traceTransaction}. All these transactions called the removed library through \texttt{DELEGATECALL}, which wasted users' gas or ETH. For malicious DoS contracts, there are 1,128 external transactions used for contract deployments. All the other transactions (15,334 external transactions and 7,700,836 internal transactions) executed with ``Out of Gas Error'' were exploited for DoS attacks.

For the 479,153 meaningless contracts, we crawl all of their 2,080 external transactions and 490,611 internal transactions for checking and debugging. Apart from 2,002 contract deployment's external transactions, 7 external transactions were halted by the first executed operation \texttt{STOP} before their data fields were processed, which verifies their uselessness. All the other 71 external transactions were executed with ``Reverted Error''. Apart from 489,890 internal transactions used for contract deployment or compulsive ETH transfer through \texttt{SELFDESTRUCT}, all the other 721 internal transactions were executed with ``Reverted Error''. 

For the 150 stack/opcode error contracts, we crawl all of their 141 external transactions and 157,513 internal transactions for analysis. Apart from 150 transactions used for contract deployment and 337 transactions used for compulsive ETH transfer through \texttt{SELFDESTRUCT}, all the other 157,167 transactions were encountered ``Bad Instruction Error'' or ``Stack Underflow Error''.

For the 195 empty accounts, we crawl all of their 237 external transactions and 5 internal transactions. Apart from 195 contract deployment's transactions, we analyze other 47 transactions. All of these 47 transactions transferred ETH or called the empty accounts with function signatures in their data fields, which denotes that they were intended to call a function of contract. However, all of their data fields were not processed because the accounts were empty, which denotes their uselessness. For the 27,200 DoS EOAs, we crawl all of their 1,163,763 internal transactions, and there does not exist external transaction. In 27,200 internal transactions, the attacker created DoS EOA through transferring 1 Wei, which is the smallest financial cost for the attacker. All the other 1,136,563 internal transactions were executed with ``Out of Gas Error'', which were used for DoS attack (analyzed in Section~\ref{sec:erasable-eoa}).

\noindent\underline{\textbf{Answer to RQ2 (Accuracy):}} All the detected erasable accounts' related transactions are useless, and more than 99.9\% of the detected erasable accounts are still stored in Ethereum.

\subsection{RQ3 Waste}
In this section, we evaluate the money lost due to erasable accounts. We analyze the gas and ETH consumed in erasable accounts' transactions, whose statistics are shown in Table~\ref{sta_accu}. For DoS contracts, 733,583,247 gas were consumed during calling Parity wallets before they were attacked. Therefore, these gas are not wasted. We analyze all the DoS contracts' balance values and 515,035.16 ETH transferred to them are permanently locked in DoS contracts, which are wasted. For meaningless contracts, all their consumed gas are wasted. However, 272.77 ETH attached to their transactions were returned to users due to ``Reverted Error'', which are not wasted. For category \ding{184} to \ding{186}, all their gas and ETH are wasted. According to the gas prices set in transactions and ETH price (204.36\$/ETH) on May 25 of 2020~\cite{14}, 106,360,910\$ is totally wasted due to these erasable accounts. 

\noindent\underline{\textbf{Answer to RQ3 (Waste):}} About 89 billion gas and 515,037 ETH are wasted due to erasable accounts, which are worth 106,360,910\$.

\section{Graph Analysis}
\label{sec:graph}
We analyze attacks/behaviors related to discovered erasable accounts to answer the question: \underline{How are erasable accounts behaved and created in reality?}

\textsc{Glaser}'s graph analysis module can be divided into two parts, i.e., call graph and creation graph. First, through symbolic execution, we analyze DoS contract's runtime bytecodes to generate call graph from erasable accounts to other accounts. According to the definitions (in Section~\ref{sec:account}) of erasable accounts, only DoS contracts can call other accounts. During symbolic execution, we analyze the operands of DoS related operations (in Section~\ref{sec:glaser}). If the target address of one operation is a real value, we can conclude that the DoS contract interacts with another account and we add an edge into the call graph. Second, through transaction analysis, we generate creation graph for erasable accounts. We analyze the account creation related transactions of erasable accounts and filter out their source addresses, which constructs the nodes of creation graph. The creation related transactions construct the edges of creation graph. If the erasable account is created through contract account, we also analyze which user (i.e., EOA) calls the contract. Furthermore, we also analyze the creation source address's transactions to see whether it creates other accounts.

\begin{figure}[ht]
	\centering
	\vspace*{-5ex}
	\includegraphics[width=4.50in]{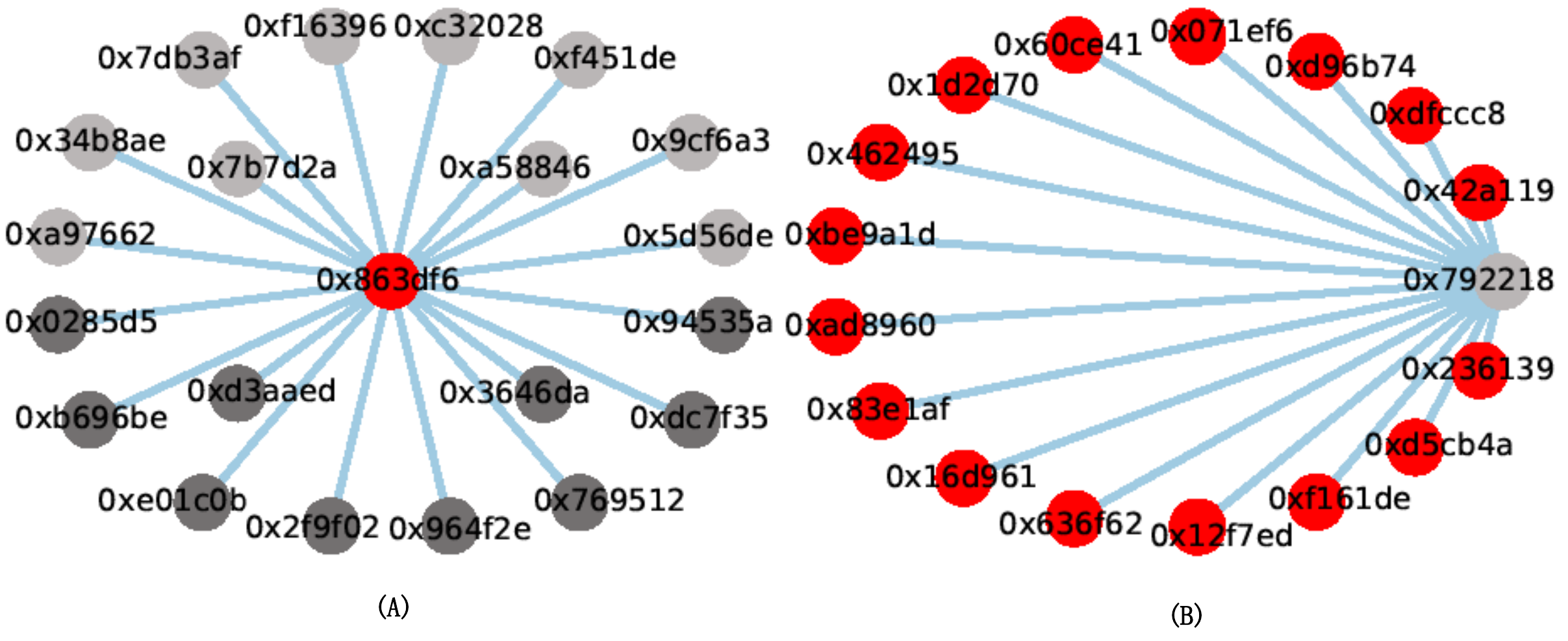}
	\vspace*{-3ex}
	\caption{(A): Call graph of Many-to-One DoS contracts (best viewed in color). The center contract has been removed (in red color), and some of its dependency error contracts (in deep grey) still have ETH balance. (B): Call graph of One-to-Many DoS contract. The malicious contract (in grey) executes massive \texttt{EXTCODESIZE} to external contracts, which have all been removed.}
	\vspace{-4ex}
	\label{character_1}
\end{figure}

\textbf{Call graph:} According to their features, the DoS contracts can be divided into two types, i.e., Many-to-One DoS contract and One-to-Many DoS contract, whose topology graphs are shown in Figure~\ref{character_1}. We only show the first three bytes of contracts' addresses for better display. 

For Many-to-One DoS contract, one center contract's address is hardcoded and interacted with many other contracts. Some Many-to-One DoS contracts detected through \textsc{Glaser} are shown in Figure~\ref{character_1} (A). In this example, the center contract (Address: \href{https://etherscan.io/address/0x863df6bfa4469f3ead0be8f9f2aae51c91a907b4}{\seqsplit{0x863D..}}) is Parity's multi-sig library, which was attacked in 2017. The center contract is removed, all its dependant wallet contracts become out of service (i.e., dependency error). \textsc{Glaser} has discovered 658 contract accounts calling the removed library. We only show 20 attacked Parity contracts in the figure for better display, and the nodes in deep grey color represent that these erasable contracts still store ETH. 

For One-to-Many DoS contract, one DoS contract hardcodes and interacts with many other contracts. One example of One-to-Many DoS contract detected through \textsc{Glaser} is shown in Figure~\ref{character_1} (B). In this example, \textsc{Glaser} has discovered that one malicious DoS contract (Address: \href{https://etherscan.io/address/0x792218d8bbe00fb81296236b014Fb14af2DA385B}{\seqsplit{0x7922..}}) hardcodes and interacts with 200 different external contracts, which have all been removed. We only show 16 removed contracts in the figure, and the malicious DoS contract (in light grey) is still stored in StateDB. Both types of DoS contracts might be called, which will result in waste of gas or ETH. For example, one DoS contract (Address: \href{https://etherscan.io/address/0x41849f3bd33ced4a21c73fddd4a595e22a3c2251}{\seqsplit{0x4184..}}) shown in Figure~\ref{character_1} has been transferred ETH in 57 transactions, which can be avoided if its account was detected/alerted in time.

\textbf{Creation graph:} According to their features, the creation graphs can be divided into two types: erasable account created by EOA, and erasable account created by contract.

\begin{figure}[ht]
	\centering
	\vspace*{-7ex}
	\includegraphics[width=4.50in]{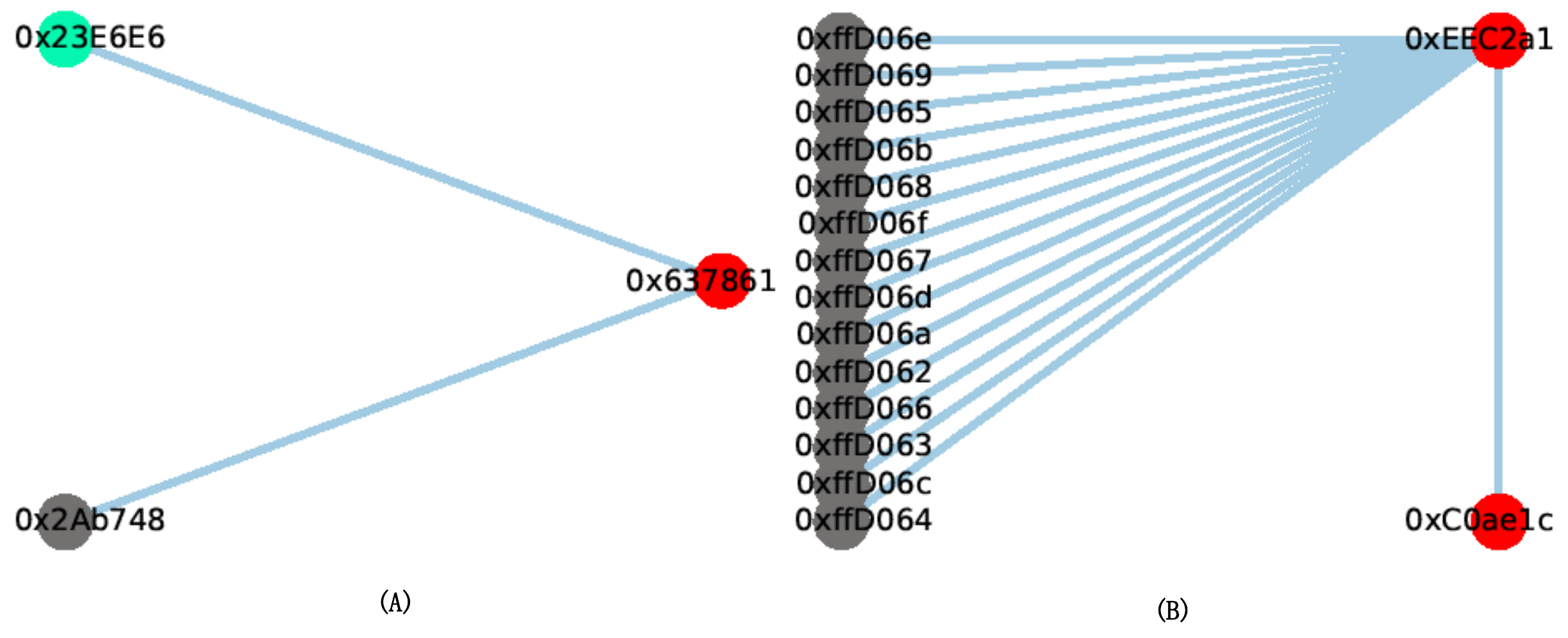}
	\vspace*{-3ex}
	\caption{(A): Creation graph of one meaningless contract. After the user (in red) realized the uselessness of the deployed contract (in grey), he redeployed another correct contract (in green). (B): Creation graph of DoS EOAs. The attacker (in red) created 14 different DoS EOAs in one transaction. We show the last three bytes of the DoS EOAs' addresses.}
	\vspace{-7ex}
	\label{character_2}
\end{figure}

Erasable accounts were created by EOAs due to programming error or deployment error, and we explain their creations through one meaningless contract example, whose creation graph is shown in Figure~\ref{character_2} (A). The user (in red color) called one deployed contract (Address: \href{https://etherscan.io/address/0x2Ab748a546760b1EC834E164DEDE2E71C4010E1d}{\seqsplit{0x2Ab7..}}) and realized that it was meaningless due to deployment error. The user incorrectly used runtime bytecodes to deploy the contract and transferred ETH to it. Then the user redeployed another correct contract (in green color), whose runtime bytecodes are just same with the data field of the transaction deploying the previous meaningless contract. Note that these types of erasable accounts' creation can be avoided, and it is better to first test and deploy contracts in private/public Testnet before they are deployed in Mainnet.

Erasable accounts were created by contracts due to some attacks, and we explain their creations through one DoS EOA example, whose creation graph is shown in Figure~\ref{character_2} (B). The attacker exploited one EOA (Address: \href{https://etherscan.io/address/0xc0ae1ca3d89a417cbe525498a1a20d40c9fd720d}{\seqsplit{0xc0ae..}}) to call a malicious contract (Address: \href{https://etherscan.io/address/0xeec2a1ee6ee942596b6e255d24d38c0a9338cfef}{\seqsplit{0xeec2..}}), creating 14 different DoS EOAs through internal transactions (Hash: \href{https://etherscan.io/tx/0xefc6cc36a06eb6b067a35e028a2ad42617d16ff9e958e8fdaec599d474e306f2}{\seqsplit{0xefc6..}}). Exploiting one storage variable, the malicious contract can generate different addresses in different transactions. These addresses were calculated and generated in runtime bytecodes, and only the last three bytes of them are different. The attacker totally created 12,204 different DoS EOAs leveraging this malicious contract, resulting in the waste of system resources and nodes' difficulty in syncing data. 


\section{Related Work}
\label{sec:relatedwork}

Frowis et al.~\cite{frowis2017code} constructed call graph for smart contracts deployed in Ethereum and discovered contracts calling to removed contracts. Note that they focus on measuring the control flow immutability between contracts, whose purpose is different from our work. Kiffer et al.~\cite{kiffer2018analyzing} measured smart contracts' creation and interaction with each other, which interpreted how are smart contracts being used. However, they do not analyze erasable contracts that exist in Ethereum. There are some other research leveraging symbolic execution~\cite{chang2018scompile}, static analysis~\cite{brent2018vandal}~\cite{chen2018towards}, and formal methods~\cite{sergey2018temporal} to analyze different kinds of security issues for smart contracts. Kiffer et al.~\cite{kiffer2018analyzing} measured the overall usage of Ethereum, which interpreted how is Ethereum being used. They discovered that \texttt{SELFDESTRUCT}'s usage rose sharply during DoS attacks in 2016. However, they do not measure or analyze erasable accounts produced during DoS attacks. Chen et al.~\cite{chen2017adaptive} proposed an adaptive gas cost mechanism for Ethereum to defend against under-priced DoS attacks. They do not analyze real accounts in Ethereum that are related to these attacks. Wang et al.~\cite{wang2018forkbase} proposed an optimization storage engine to reduce nodes' storage volume, which can improve the scalability of blockchain systems. They do not analyze the erasable accounts which are already stored in StateDB. Angelo et al.~\cite{di2019collateral} analyzed contract deployment code patterns which were exploited by attackers, and they described three related attack scenarios in reality appeared in the middle of 2018, whose contents and purposes are different from ours. They focus on the vulnerabilities and attacks leveraging skillfully crafted deployment codes, while we detect erasable accounts due to programming or deployment errors.

\section{Discussion and Conclusion}

\textbf{Discussion:} We discuss validity threats, limitations, and future work. \underline{(1)} \textsc{Glaser} can not only discover erasable accounts that already exist in Ethereum, but also erasable accounts that might be created in future. Some kinds of accounts analyzed by \textsc{Glaser} might also be created in future, and \textsc{Glaser} might discover more erasable accounts. \underline{(2)} For the discovered erasable accounts, only part of meaningless contracts can be destructed by ordinary users. Because some MC-RS have \texttt{SELFDESTRUCT} in their first basic blocks, which can be invoked through transactions by users. Although most of discovered erasable accounts cannot be easily destructed by users, our results can remind users not to call them, which can help users save money. \underline{(3)} Path explosion and timeout exception are common threats for the symbolic execution techniques leveraged in this paper. However, we use some methods to reduce these threats. During detecting stack error contracts, we first extract runtime bytecodes corresponding to the first basic block and then symbolically execute them. During detecting attacked Parity wallets, we first filter out contracts without external call operations and then symbolically execute them. \underline{(4)} As \textsc{Glaser} focuses on five kinds of erasable accounts in Ethereum, we will detect more kinds of erasable accounts in future. We will also analyze erasable accounts in other blockchain systems.

\textbf{Conclusion:} We have conducted the first work that systematically characterizes erasable accounts in Ethereum, i.e., erasable contract accounts and erasable EOAs. We have implemented \textsc{Glaser} to analyze the StateDB, which can detect erasable accounts leveraging bytecodes' static analysis, symbolic execution, transaction analysis, and state fields analysis. Furthermore, we have analyzed attacks/behaviors related to erasable accounts through graph analysis. Extensive experiments are also conducted to evaluate the quantity, accuracy, and waste of the detected erasable accounts.


\bibliographystyle{splncs04}
\bibliography{ref}

\end{document}